\documentclass [prl,twocolumn,superscriptaddress,floatfix,amsmath,amssymb]{revtex4-2}
\usepackage{amsmath}
\usepackage{graphicx}
\usepackage{hyperref}
\usepackage{color}
\usepackage{amssymb}
\usepackage{bm}
\usepackage[latin9]{inputenc}
\usepackage{mathtools}
\usepackage{eufrak}
\usepackage[normalem]{ulem}
\usepackage{subfigure}
\makeatletter
\usepackage{comment}
\newcommand{\beq}{\begin{equation}}
\newcommand{\eeq}{\end{equation}}
\newcommand{\bea}{\begin{eqnarray}}
\newcommand{\eea}{\end{eqnarray}}

\@ifundefined{textcolor}{}{%
 \definecolor{BLACK}{gray}{0}
 \definecolor{WHITE}{gray}{1}
 \definecolor{RED}{rgb}{1,0,0}
 \definecolor{GREEN}{rgb}{0,1,0}
 \definecolor{BLUE}{rgb}{0,0,1}
 \definecolor{CYAN}{cmyk}{1,0,0,0}
 \definecolor{MAGENTA}{cmyk}{0,1,0,0}
 \definecolor{YELLOW}{cmyk}{0,0,1,0}
}

\usepackage{epstopdf}
\usepackage{color}
\usepackage{array}
\usepackage{mathrsfs}
\usepackage{dcolumn}% Align table columns on decimal point
\usepackage{bm}% bold math
\usepackage{hyperref}

\newcolumntype{L}[1]{>{\raggedright\let\newline\\\arraybackslash\hspace{0pt}}m{#1}}
\newcolumntype{C}[1]{>{\centering\let\newline\\\arraybackslash\hspace{0pt}}m{#1}}
\newcolumntype{R}[1]{>{\raggedleft\let\newline\\\arraybackslash\hspace{0pt}}m{#1}}

\graphicspath{{../Inkscape/},{../Mathematica/},{../Mathematica/Paper_figures/},{../Matlab/}}

\usepackage{tikz}
\usetikzlibrary{calc,intersections}
\usetikzlibrary{shadings}
\usepgflibrary{shadings}

\makeatother

\begin{document}
\title{Dynamical vortices in electron-phonon superconductors}
\author{Morten H. Christensen}
\email{mchriste@nbi.ku.dk}
\affiliation{School of Physics and Astronomy, University of Minnesota, Minneapolis,
	Minnesota 55455, USA}
\affiliation{Center for Quantum Devices, Niels Bohr Institute, University of Copenhagen, 2100 Copenhagen, Denmark}
\author{Andrey V. Chubukov}
\affiliation{School of Physics and Astronomy, University of Minnesota, Minneapolis,
Minnesota 55455, USA}
\date{\today}
\begin{abstract}
We analyze the structure of an $s-$wave superconducting gap in systems with electron-phonon attraction and electron-electron repulsion. Earlier works have found that superconductivity develops despite strong repulsion, but the gap, $\Delta (\omega_m)$,  necessarily changes sign along the Matsubara axis. We analyze the sign-changing gap function from a topological perspective using the knowledge that a nodal point of $\Delta (\omega_m)$ is the center of dynamical vortex. We consider two models with different cutoffs for the repulsive interaction and trace the vortex positions along the Matsubara axis and in the upper frequency half plane upon changing the relative strength of the attractive and repulsive parts of the interaction. We discuss how the presence of dynamical vortices affects the gap structure along the real axis, detectable in ARPES experiments.
\end{abstract}
\maketitle

{\bf Introduction.--}In recent years there has been a tremendous amount of interest in fundamental properties of superconductors that go beyond breaking of the $U(1)$ phase symmetry. Possible candidates are superconductors that additionally break either time-reversal symmetry~\cite{nandkishore2012chiral,sigrist,Fu}, or lattice-rotational symmetry (nematic superconductors)~\cite{kushnirenko2020nematic,graphene_nematic}. Another option is topological superconductivity, either induced by proximity to a topological insulator, or developing on its own~\cite{Fu,hasan,ando}. Simultaneously, there have been multiple studies in recent years of how superconductivity emerges from a nominally repulsive interaction~\cite{scalapino,acs,review2,review3,paper_1}. The dominant theme of this research is the analysis of how superconductivity with a spatial gap structure different from an ordinary $s-$wave emerges due to screening of a bare repulsion by the Kohn-Luttinger mechanism, extended to lattice systems and to cases where screening arises from soft collective excitations in the spin or charge channel~\cite{review4}. At the same time, renewed interest in superconductivity in SrTiO$_3$ and other low-density materials~\cite{Behnia_2019} has triggered a re-examination of how ordinary $s-$wave superconductivity emerges in systems with strong Coulomb repulsion and weaker electron-phonon attraction~\cite{Fernandes_rev,Ruhman2016,Maria2019,Prokofiev2019}.
Here and below, by attraction and repulsion we mean the sign of a dynamical interaction $V(\omega_m-\omega_{m'})$ on the Matsubara axis, where $V$ is real and its sign is a well-defined quantity. On the real axis, a dynamical $V(\omega-\omega')$ is a complex function of frequency and there is no direct way to determine its sign.

The ``conventional'' argument for the existence of superconductivity despite the presence of stronger electronic repulsion goes as follows: The Coulomb potential is renormalized down by scattering in the particle-particle channel in the interval between the Fermi energy, $E_F$, and the Debye frequency, $\omega_D$. If the interval is wide enough, the downward renormalization is strong, and at frequencies below $\omega_D$ the Coulomb repulsion becomes smaller than the electron-phonon attraction~\cite{mcMillan,ummarino}. This reduced form of the Coulomb interaction is often referred to as the Morel-Anderson pseudopotential~\cite{Morel-Anderson} and, in dimensionless units, is denoted $\mu^*$. However, this reasoning requires care as the total dynamical four-fermion interaction -- the sum of Coulomb repulsion and electron-phonon attraction -- actually remains repulsive at all frequencies, even after it is renormalized in the particle-particle channel between $E_F$ and $\omega_D$. A more accurate analysis~(see e.g. Ref.~\onlinecite{Rietschel_1983}) shows, that the system finds a way to neutralize the overall repulsion by developing a frequency dependent $s-$wave gap, $\Delta (\omega_m)$, which changes sign between small and large frequencies. A ``conventional'' description of electron-phonon superconductivity, in which the effect of the Coulomb interaction reduces to $\mu^*$ and the gap function is nearly frequency independent, emerges only after one integrates out fermions with larger $\omega_m$ for which $\Delta (\omega_m)$ has a different sign. The same care needs to be applied to a semi-phenomenological model~\cite{,ummarino,Prokofiev2019}, in which electron-phonon and electron-electron interactions are treated separately, and the Coulomb repulsion is replaced by an on-site Hubbard $U$ term. In this model, the total interaction can be made attractive at small frequencies by adjusting the strength of the Hubbard $U$. However, if the upper cutoff for the $U$ term is much larger than the Debye frequency, the total interaction necessarily becomes repulsive at larger frequencies. Then, again, the gap function must change sign between low and high frequencies.

In this communication we analyze the sign change of $\Delta (\omega_m)$ from a topological perspective, bringing together the two directions of research on unconventional superconductivity. It has been argued recently~\cite{paper_vort} that a zero of the gap function on the Matsubara axis is a center of a dynamical vortex, around which the phase of $\Delta$ winds by $2\pi$ under an anticlockwise rotation. Superconducting states with and without a zero in $\Delta (\omega_m)$ are therefore topologically distinct, and the state with a zero is a dynamical analog of a nodal topological superconductor~\cite{ando}. Vortices on the Matsubara axis are subtly different from their real-space analogues: A vortex in real space can appear (disappear) through generation (annihilation) of a vortex-antivortex pair. For a dynamical vortex this is not possible, because a dynamical antivortex corresponds to a pole in $\Delta (\omega_m)$ and cannot emerge in the upper frequency half-plane, where $\Delta$ is analytic. Hence, a single vortex cannot just appear on the positive Matsubara axis and has to either come from infinity, or from the lower frequency half-plane. On the other hand, a pair of vortices can appear (disappear) at a given point on the positive Matsubara axis as a result of merging (splitting) of two vortices in the upper frequency half-plane (by the Cauchy relation, if there is a vortex at a complex $z=\omega' + i \omega^{''}$, there must be another one at $z=-\omega' + i \omega^{''}$).

{\bf Nodal points on the Matsubara axis as dynamical vortices.--}To elucidate these distinct situations, consider first the case when $\Delta (\omega_m)$ changes sign once at $\omega_m = \omega_0$. Near this frequency, $\Delta_0 (\omega_m) = - c (\omega_m - \omega_0)$ (we set $c >0$ for definiteness). Let us analytically continue $\Delta (\omega_m)$ to the vicinity of the Matsubara axis, i.e., to $z = \omega' + i \omega^{''}$ (on the Matsubara axis, $z = i \omega_m$).  Because $\Delta (\omega_m)$ is non-singular, $\Delta (z) = \Delta (i\omega_m \to \omega' + i \omega^{''})$. For any non-zero $\omega'$,  $\Delta (z)$ is a complex function: $\Delta (z) = \Delta' (z) + i \Delta^{''} (z)$, and we can introduce the phase of $\Delta (z)$ as $\eta (z) ={\text{Im}} [\log{\Delta (z)}]$. Evaluating $\eta (z)$, we find that it varies by $2\pi$ upon an anticlockwise circulation around $\omega_0$. This implies that the nodal point at $\omega_0$ is the center of a dynamical vortex. One can straightforwardly verify that a vortex on the Matsubara axis gives rise to a $2\pi$ phase variation on the real axis, between $-\infty$ and $\infty$. To see this, one should compute $\int_{-\infty}^\infty d \omega  \partial \eta (\omega)/\partial \omega$ by closing the integration contour in the upper half-plane. The function $\partial \eta (\omega)/\partial \omega$ has a simple pole at the nodal point, and modifying the contour to a circle around this point, one obtains a $2\pi$ variation of the phase.

Now suppose that there are two nodal points on the Matsubara axis. One can straightforwardly verify that each is the center of a vortex with a $2\pi$ anticlockwise circulation. When these two points are close to each other, at $\omega_{a,b} = \omega_0 \pm \delta$, one can approximate the gap function near these points as $\Delta (\omega_m) \approx {\bar c} (\omega_m - \omega_0 -\delta) (\omega_m - \omega_0 + \delta)$. Suppose that $\delta = \delta (x)$ varies under some parameter, vanishes as $x$ approaches some $x_0$ from below, and then becomes imaginary $\delta (x >x_0) = i {\bar \delta} (x)$. The two vortices approach each other as $x$ approaches $x_0$ from below and  merge at $\omega_0$ at $x =x_0$. However, they do not annihilate because they have the same ``charge''. Indeed, one can easily check that the anticlockwise circulation around $\omega_0$ becomes $4\pi$. At large $x$, $\Delta (\omega_m) = (\omega_m-\omega-0)^2 + {\bar \delta}^2(x)$ is now nodeless on the Matsubara axis. Extending the gap function to the upper half-plane of complex frequency, $i\omega_m \to z = a + ib$, we find that the vortices split and move to $a = \pm{\bar \delta}(x)$ and $b=\omega_0$. We illustrate this in Fig.~\ref{fig:phase_figure_hard_cutoff}.

\begin{figure}
	\includegraphics[width=0.98\columnwidth]{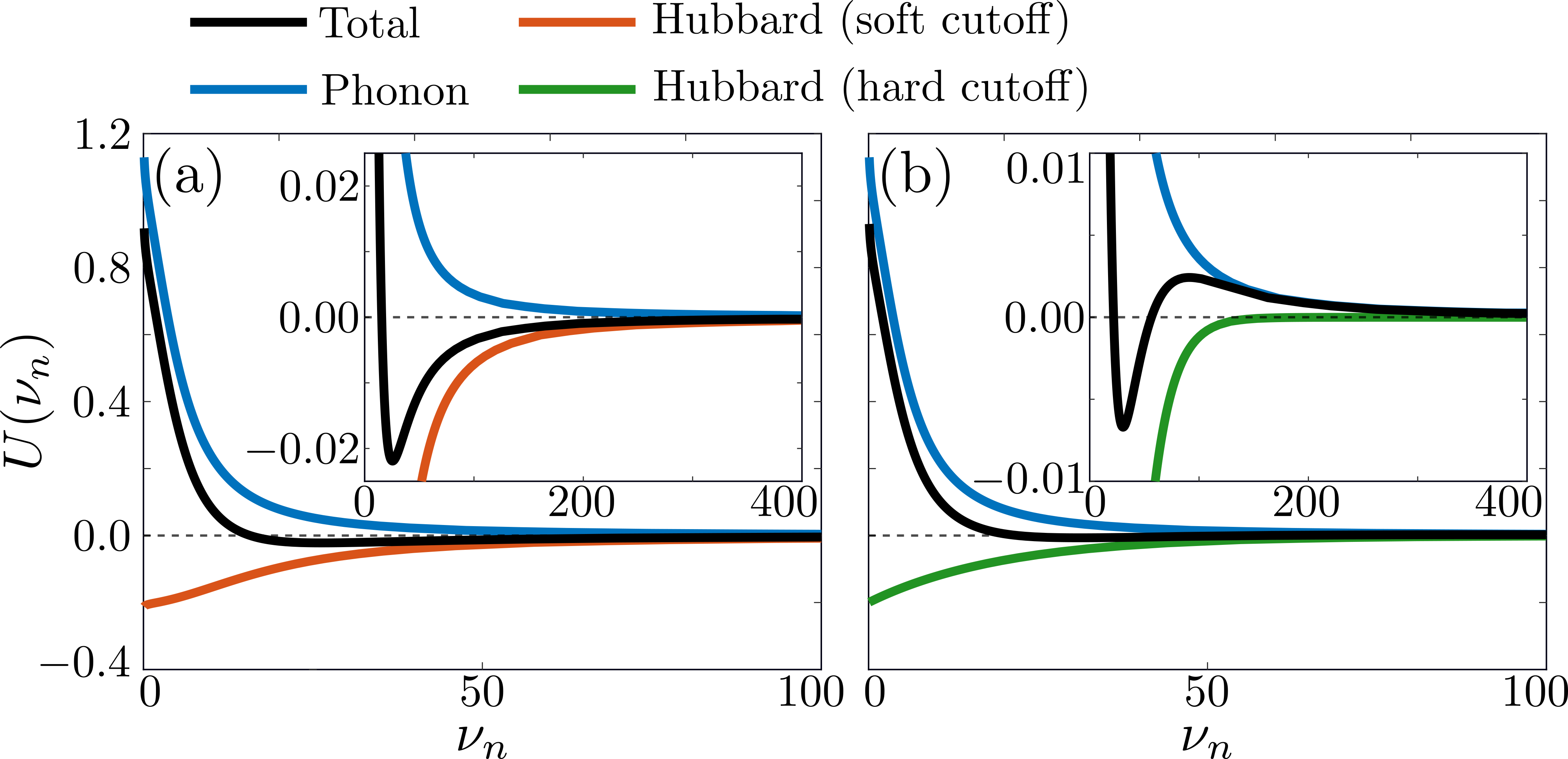}
	\caption{\label{fig:interactions} Total pairing interaction as the sum of electron-phonon attraction, $U_{\rm el-ph} (\nu_n)$, and Hubbard repulsion $U_{\rm Hub}(\nu_n)$, for which we impose either a soft (a) or hard (b) cutoff. The insets show a zoom of the $y$-axis alongside an extended view of the $x$-axis to highlight the number of sign-changes. In (a) the total interaction is repulsive at high frequencies while at low frequencies, it can be either repulsive or attractive, depending on the parameters. Here, we choose it to be attractive. In (b) the total interaction is attractive at both low and high frequencies, but can be repulsive at intermediate frequencies.}
\end{figure}

{\bf Model.--}For the actual calculations of vortex positions and trajectories, we consider two models with Hubbard repulsion and electron-phonon attraction. In both cases we will assume that the electron-phonon interaction is $U_{\rm el-ph} (\nu_n) = g \nu^2_0/(\nu_n^2 + \nu^2_0)$~\footnote{
This form for the interaction leads to singularities on the real frequency-axis~\cite{Marsiglio2020}. To smoothen these, we employ a broadened kernel and the electron-phonon interaction and the soft cutoff Hubbard term is $U(\nu_n)=\int_{0}^{\infty}\mathrm{d}\nu\frac{2\nu}{\nu^2+\nu_n^2}\frac{\lambda \Omega}{2\pi}\left[ \frac{\Gamma}{(\nu-\Omega)^2+\Gamma^2}-\frac{\Gamma}{\Omega_c^2 + \Gamma^2} \right]\theta\left(\Omega_c - |\nu-\Omega| \right)$. Here $\lambda$ is either $g$ or $U$, $\Omega$ is $\nu_0$ or $\omega_c^{\rm soft}$ and $\Omega_c$ ensures that the integral is convergent.}, with the electron-phonon coupling $g$, the phonon frequency $\nu_0$, and $\nu_n=\omega_m-\omega_{m'}$. We set the parameters such that, at small frequencies, the electron-phonon attraction is stronger than the Hubbard repulsion and focus on how vortices appear when the total interaction becomes repulsive at higher frequencies. In the first case, denoted Model I, we assume a soft cutoff for the Hubbard term, i.e., take it to be $U_{\rm Hub}(\nu_n)=U(\omega^{\rm soft}_c)^2/(\nu_n^2 + (\omega^{\rm soft}_c)^2)$. In this case, both $U_{\rm Hub}(\nu_n)$ and the electron-phonon interaction scale as  $1/\nu^2_n$ at the highest frequencies. The total interaction, $U_{\rm el-ph} (\nu_n) + U_{\rm Hub}(\nu_n)$, is then either positive at all frequencies, or undergoes a single sign-change, as illustrated in Fig.~\ref{fig:interactions}(a).
Correspondingly, $\Delta (\omega_m)$ is either sign-preserving or has a single zero. We show that, in this situation, a vortex emerges at $\omega_m = \infty$ at some critical $\nu_0$ and moves down along Matsubara axis as $\nu_0$ is reduced. In the second case, denoted Model II, we use a hard cutoff for the Hubbard repulsion, i.e. assume that it decays exponentially above a certain scale $\omega_c^{\rm hard}$, $U_{\rm Hub}(\nu_n)=U \exp ( -|\nu_n|/\omega_c^{\rm hard} )$. The total interaction is then attractive at both small and large frequencies, but for sufficiently small $\nu_0$ (at a given $U$ and $\omega_c^{\rm hard}$) it becomes repulsive at intermediate frequencies, as illustrated in Fig.~\ref{fig:interactions}(b). In this case $\Delta (\omega_m)$ has the same sign at small and large $\omega_m$, but when $\nu_0$ is smaller than some critical value, it has two zeros on the Matsubara axis at finite $\omega_m$. As we show below, as $\nu_0$ is reduced, a pair of vortices first appears on the real axis, upon crossing from the lower to the upper frequency half-plane, and then moves towards the Matsubara axis. The two vortices merge on the Matsubara axis at a critical value of $\nu_0$, at which point $\Delta (\omega_m)$ develops a double zero at some $\omega_m = \omega_0$, and then split along the Matsubara axis, leading to two sign changes of $\Delta (\omega_m)$.

 \begin{figure}
\includegraphics[width=0.98\columnwidth]{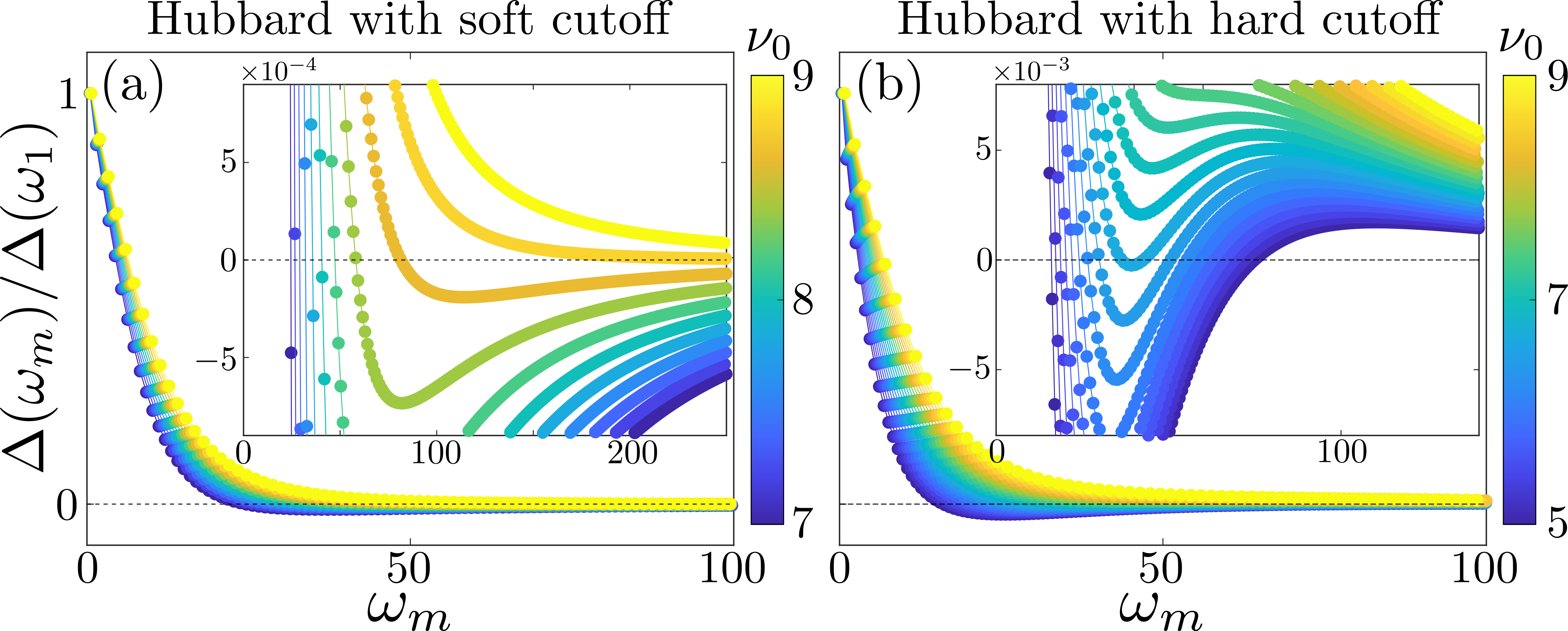}
\caption{\label{fig:matsubara_gap}Gap function along Matsubara axis, $\Delta (\omega_m)$, for (a) Model I a with soft cutoff and (b) Model II with a hard cutoff. The insets are zoom-ins near $\Delta(\omega_m)=0$ and the color denotes $\nu_0$.  In (a) the gap is sign-preserving for larger $\nu_0$ and changes sign once for smaller $\nu_0$. In (b), the sign of the gap at high and low frequencies is the same. The full $\Delta (\omega_m)$ is either sign-preserving, or changes sign twice.}
\end{figure}

 \begin{figure*}
\includegraphics[width=0.98\textwidth]{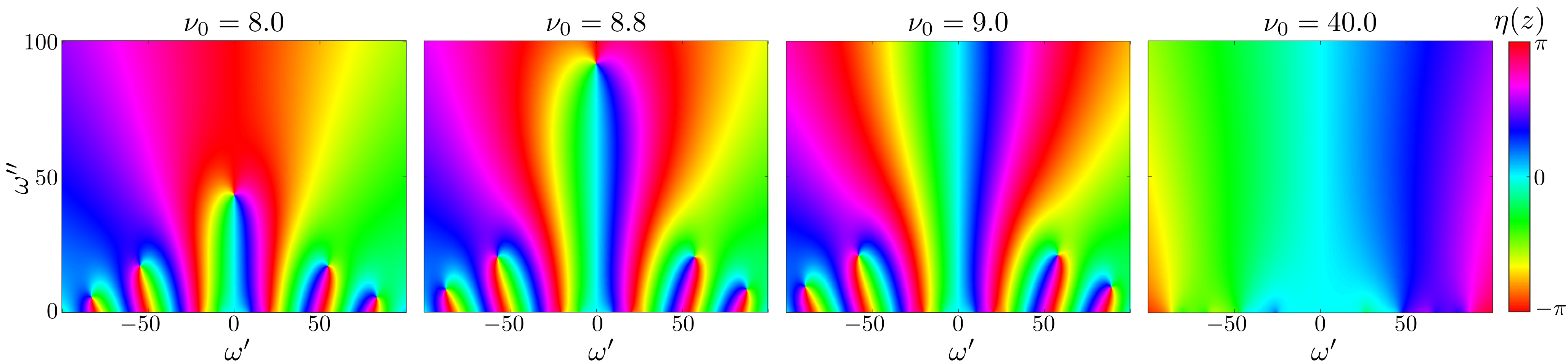}
\caption{\label{fig:phase_figure_soft_cutoff}The phase of the gap function, $\eta (z)$, in the upper half-plane of frequency for Model I with a soft cutoff for the Hubbard repulsion [see Fig.~\ref{fig:matsubara_gap}(a)], where $\Delta (z) = |\Delta (z)| e^{i\eta (z)}$, $z = \omega' + i \omega^{''}$. Along the Matsubara axis, $z = i \omega_m$. The periodic color scale denotes the value of the phase and the points where all the colors meet correspond to vortex locations, around which the phase varies by a factor of $2\pi$.}	
\end{figure*}
 \begin{figure*}
\includegraphics[width=0.98\textwidth]{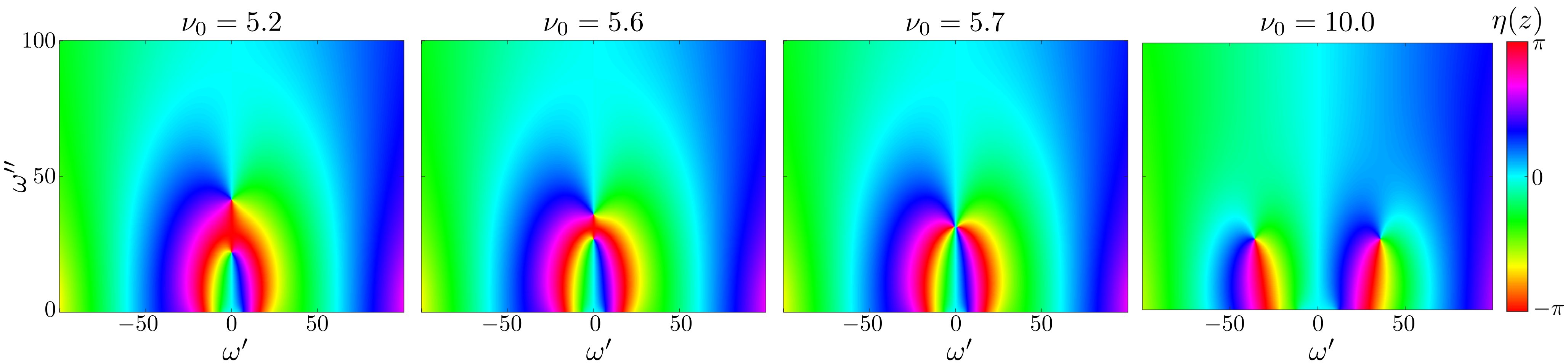}
\caption{\label{fig:phase_figure_hard_cutoff}The same as in Fig.~\ref{fig:phase_figure_soft_cutoff}, but for Model II with a hard cutoff for the Hubbard repulsion [see Fig.~\ref{fig:matsubara_gap}(b)]. The vortices meet on the Matsubara axis for $\nu_0 \approx 5.7$ and split off into the complex frequency plane. Note that, encircling the point where the vortices meet near $\omega''\approx 30$ yields a phase of $4\pi$, as explained in the text.}
\end{figure*}

\begin{figure}
\includegraphics[width=0.98\columnwidth]{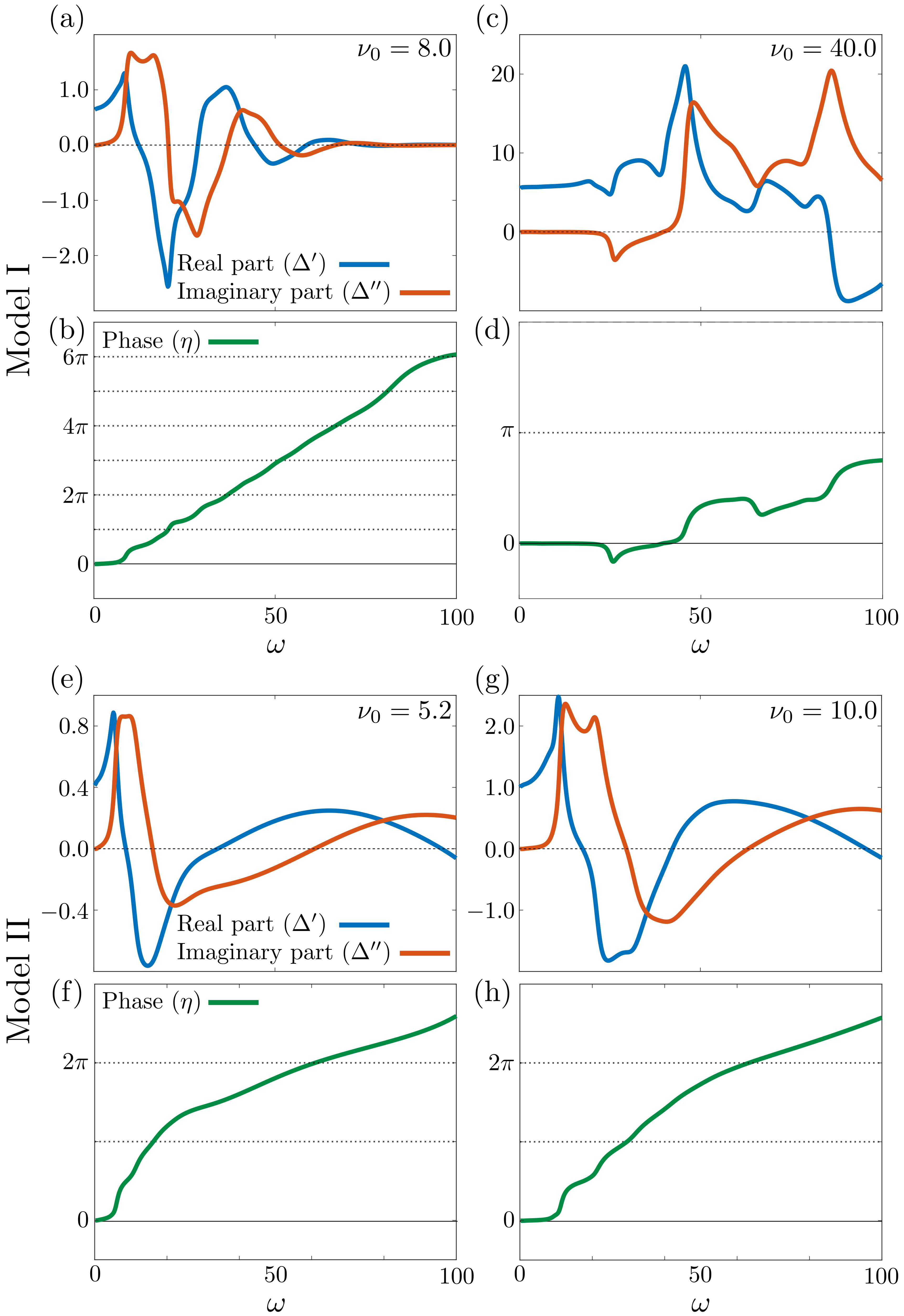}
\caption{\label{fig:gap_real_axis}The behavior of the gap function and its phase along the real axis for (a)-(d) Model I and (e)-(h) Model II. The gap $\Delta(\omega) = \Delta' (\omega) + i \Delta'' (\omega)$ is complex and both $ \Delta' (\omega)$ (blue) and $\Delta'' (\omega)$ (red) change sign multiple times for smaller $\nu_0$. The total change of the phase between $\omega = - \infty$ and $\omega = \infty$ is $\delta \eta = 2\pi (n+1)$, where $n$ is the number of vortices in the upper half-plane of frequency. This is reflected in the phase variation, shown in (b), (d), (f), and (h).}
\end{figure}

{\it \bf  Gap function on the Matsubara axis.--}We obtain the gap function within Eliashberg theory, by solving the integral gap equation
\begin{equation}
   \Delta (\omega_m) = \pi T \sum_{m'}  \frac{\Delta (\omega_{m'}) - \Delta (\omega_m) \frac{\omega_{m'}}{\omega_m}}{\sqrt{(\omega_{m'})^2 +\Delta^2 (\omega_{m'})}} U_{\rm tot} \left(\omega_m - \omega_{m'}\right)\,.
     \label{eq:1}
\end{equation}
with the dynamical interaction $U_{\rm tot} (\nu_n) = U_{\rm el-ph}(\nu_n) + U_{\rm Hub}(\nu_n)$. We are interested in the solution of the non-linear gap equation inside the superconducting phase and below we show results for $T/T_c=0.5$. We measure all energy variables in units of $g$ and set $U=0.2$, and $\omega_{c}^{\rm soft}=\omega_c^{\rm hard}=20$. The computational procedure has been discussed extensively, see e.g. Ref.~\onlinecite{Marsiglio2020}, and we will focus on the results.

At small frequencies $\omega_m \sim T$, relevant $\omega_{m'}$ are of the order of $\omega_m$, and $U_{\rm tot}(\nu_n) \approx g - U >0$ by assumption. The solution of the gap equation in this frequency range yields a regular, sign-preserving $\Delta (\omega_m)$, which tends to a finite value at the smallest $|\omega_m| = \pi T$ and decreases at larger $\omega_m$. In the opposite limit of high frequencies, a simple examination of Eq.~\eqref{eq:1} shows that the gap function decays as $1/\omega^2_m$. For such $\omega_m$, one can neglect $\Delta (\omega_m)$ in the r.h.s. of Eq.~\eqref{eq:1} and pull out $U_{\rm tot} (\omega_m)$ from the summand. This yields the relation
\begin{equation}
   \Delta (\omega_m) = A U_{\rm tot} (\omega_m),~~ A = \pi T \sum_{m'}  \frac{\Delta (\omega_{m'})}{\sqrt{(\omega_{m'})^2 +\Delta^2 (\omega_{m'})}}\,.
     \label{eq:2}
\end{equation}
The sum in the r.h.s. converges at large $m'$, hence $A$ is finite. This  justifies pulling $U_{\rm tot} (\omega_m)$ from the summand in  (\ref{eq:1}).

For Model I, at large frequencies, $\omega_m \gg \nu_0,\omega_1$, we have
$U_{\rm tot} (\omega_m) = (g \nu^2_0 - U \omega^2_1)/|\omega^2_m|$. The prefactor changes sign between $\nu_0 > \nu_c = \omega_1 \sqrt{U/g}$ and $\nu_0 < \nu_c$. For $\nu_0 > \nu_c$, the sign of $\Delta (\omega_m)$  in Eq.~\eqref{eq:2} is the same as that of $A$ in the summand in Eq.~\eqref{eq:2}, and it is natural to assume that $\Delta (\omega_m)$ is sign-preserving. For $\nu_0 < \nu_c$, the prefactor is negative, hence $\Delta (\omega_m)$ must change sign. We show the numerical results for Model I in Fig.~\ref{fig:matsubara_gap}(a). Indeed, $\Delta (\omega_m)$ changes sign at some frequency when $\nu_0 < \nu_c$.

For Model II, $U_{\rm Hub}(\omega_m)$ vanishes exponentially at frequencies $\omega_m \gg \omega_1$, and hence $U_{\rm tot} (\omega_m)$ is necessarily positive. In this situation, the sign of $\Delta (\omega_m)$ is the same as the sign of $A$ in (\ref{eq:2}). The latter is determined by $\Delta (\omega_m)$ at the smallest $\omega_m \sim T$, where the gap function is the largest. As a consequence, the sign of $\Delta(\omega_m)$ at small and large $\omega_m$ is the same. The full $\Delta (\omega_m)$ then either remains sign-preserving or changes sign an even number of times, most realistically twice. We show the numerical results for Model II in Fig.~\ref{fig:matsubara_gap}(b). We see that, as expected, the gap function is either sign-preserving or changes sign twice, depending on the value of $\nu_0$,

{\it \bf Dynamical vortices in the complex plane.--}Extending the analysis to complex frequencies $z= \omega' + i \omega^{''}$ we obtain the gap function $\Delta(z)$ in the entire upper complex frequency plane. This will allow us to see how a pair of vortices moves away from the Matsubara axis  in Model II and also to check whether there are additional vortices away from the Matsubara axis. Our rational for searching for these extra vortices is the study in Refs. \onlinecite{paper_vort,Paper_5}, which showed that a set of vortices near the real axis necessarily develops in the special limit when $U=0$, $\nu_0 \to 0$ but $g \nu^2_0$ tends to a constant.

The gap function $\Delta (z)$ in the upper half-plane is obtained through a two-step procedure: First, the gap function is analytically continued to the real axis, $i\omega_m \rightarrow \omega+i 0^+$, through an iterative procedure described in Ref.~\onlinecite{Marsiglio1988}. Second, the gap function on the real axis, $\Delta(\omega+ i0^+)=\Delta'(\omega+i0^+)+i\Delta''(\omega+i0^+)$, is extended to the upper complex plane using Cauchy's formula:
\begin{equation}
	\Delta(z) = \frac{2}{\pi}\int_0^{\infty}\frac{\bar{\omega}\Delta''(\bar{\omega}+i0^+)}{\bar{\omega}^2 - z^2}\mathrm{d}\bar{\omega}
\label{eq:KK}
\end{equation}
where $z=\omega'+i\omega''$. We note that, due to numerical inaccuracy, the gap function, obtained from Eq.~\eqref{eq:KK} along the imaginary axis, and the original $\Delta(\omega_m)$ turn out to be slightly different. This does not affect our results, and only leads to small variations in $\nu_c$ between the one extracted from $\Delta(\omega_m)$ and the one from $\Delta(z)$ (compare e.g. Figs.~\ref{fig:matsubara_gap} and \ref{fig:phase_figure_soft_cutoff}). We show the results for the phase of the gap function, $\eta (z)$, in the upper frequency half-plane in Fig.~\ref{fig:phase_figure_soft_cutoff} for Model I and in Fig.~\ref{fig:phase_figure_hard_cutoff} for Model II for a range of representative values of $\nu_0$.

For Model I, we see from Fig.~\ref{fig:phase_figure_soft_cutoff} that for the largest $\nu_0 =40.0$, where we expect the gap to be sign-preserving along the Matsubara axis, there are no vortices in the upper half-plane of complex frequency. For smaller $\nu_0$, a vortex appears on the Matsubara axis, located where $\Delta (\omega_m)$ in Fig.~\ref{fig:matsubara_gap}(a) changes sign. As we anticipated, the vortex appears first at $\omega_m=\infty$ and moves towards smaller $\omega_m$ as $\nu_0$ decreases.  Interestingly, we found additional vortices in the upper half-plane at smaller $\omega^{''}$ and larger $\omega'$. These additional vortices move into the upper frequency half-plane from the lower one, as $\nu_0$ is decreased. This is again consistent with our expectation.

For Model II, we see from Fig.~\ref{fig:phase_figure_hard_cutoff} that, for smaller $\nu_0$, there are two vortices on the Matsubata axis. As $\nu_0$  increases, the distance between the vortices shrinks, and once $\nu_0$ exceeds some critical value, the two vortices leave the Matsubara axis and move in opposite directions in the upper half-plane of complex frequency, gradually approaching the real axis. This is consistent with the analytical treatment earlier in the paper.

{\it \bf Gap along the real axis.--}Further evidence for the presence of the vortices both on the Matsubara axis and away from it comes from the analysis of the complex gap function along the real axis. We show $\Delta' (\omega)$ and $\Delta^{''} (\omega)$ in Fig.~\ref{fig:gap_real_axis} for both Model I and II. Fig.~\ref{fig:gap_real_axis} also shows the variation of the phase $\eta (\omega)$ of $\Delta (\omega + i0^+) = |\Delta (\omega)| e^{i\eta (\omega)}$. In simple terms, $\eta (\omega)$ increases by $\pi/2$ in each frequency interval between points where first $\Delta' (\omega)$ and then $\Delta''(\omega)$ changes sign [compare e.g. Fig.~\ref{fig:gap_real_axis}(a) and (b)].

The relation between the phase shift on the real axis and the number of vortices in the complex plane can be found~\cite{paper_vort} by evaluating $\delta \eta = \int_{-\Omega}^\Omega d \omega  \partial \eta (\omega)/\partial \omega$ by closing the integration contour over the upper half-plane. The function $\partial \eta (\omega)/\partial \omega$ is analytic in the upper half-plane except at the nodal points of the gap function, where it has simple poles. Evaluating the integral, one obtains  $\delta \eta = \delta \eta_\Omega+ 2\pi n$, where $n$ is the number of vortices in the upper half-plane and $\delta \eta_\Omega$ is the phase shift without vortices. The latter is $2\pi$ at $\Omega \to \infty$ because, at the largest $\omega$,  $\Delta (\omega) \propto
e^{i \pi \text{sgn} \omega}$, but is generally smaller  when $\Omega$ is finite, and its value is  model-dependent. The frequency interval shown in Figs.~\ref{fig:phase_figure_soft_cutoff} and \ref{fig:phase_figure_hard_cutoff} corresponds to $\Omega=100$ (recall that all frequencies are in units of $g$).

For Model I, let us first consider the case where $\nu_0=40.0$. In this case, there are no vortices and $\delta\eta = \delta \eta_{\Omega}$. From the phase variation shown in Fig.~\ref{fig:phase_figure_soft_cutoff}, we see that $\delta \eta_\Omega \lesssim 2\pi$. This matches the variation of the phase shown in Fig.~\ref{fig:gap_real_axis}(d) in the interval between 0 and $\Omega$, corresponding to positive $\omega'$ in Fig.~\ref{fig:phase_figure_soft_cutoff}, which is slightly less than $\pi$. For smaller $\nu_0=8.0$, $\delta \eta_{\Omega}\lesssim 2\pi$ (blue on the far left, green on the far right in Fig.~\ref{fig:phase_figure_soft_cutoff}), and there are five vortices. Hence, the total phase variation, $\delta\eta\approx 12\pi$, implying a $6\pi$ variation in the interval between $0$ and $\Omega$. This matches the result shown in Fig.~\ref{fig:gap_real_axis}(b).

For Model II, we see from Fig.~\ref{fig:phase_figure_hard_cutoff} that for all values of $\nu_0$, $\delta \eta_\Omega \gtrsim \pi$, and there are two vortices either on the Matsubara axis, or away from it (for larger $\nu_0$). Then, $\delta \eta \approx 5\pi$, corresponding to $5\pi/2$ between $\omega =0$ and $\omega =\Omega$. Reading off $\delta \eta$ from Fig.~\ref{fig:gap_real_axis}(f) and (h), we find the same number.

{\it \bf Conclusions.--}In this work we analyzed the structure of an $s-$wave superconducting gap $\Delta (\omega_m)$ in systems with frequency-dependent electron-phonon attraction and electron-electron repulsion, which we approximated by Hubbard $U$.  Previous works have found that superconductivity develops even when electron-electron repulsion is larger, but then the gap necessarily changes sign along the Matsubara axis. We analyzed the sign-changing  gap function from a topological perspective, by taking advantage of the knowledge that a nodal point of $\Delta (\omega_m)$ is the center of a dynamical vortex. To this end,  we considered two models with soft and hard high-frequency cutoffs for the for the  Hubbard $U$ term. In both models, we assumed that the total interaction is attractive at small frequencies and used the phonon frequency $\nu_0$ as the knob to calibrate the relative strength of the attractive and repulsive components of the interaction at high frequencies. In Model I with a soft cutoff, a vortex first emerges at $\omega_m = \infty$ upon the reduction of $\nu_0$, and moves down along the Matsubara axis. Simultaneously, additional vortices cross from the lower half-plane of frequency into the upper one, but remain away from  the Matsubara axis. For Model II with a hard cutoff, we found that, as $\nu_0$ is reduced, two vortices cross from the lower frequency half-plane into the upper one, and start moving towards the Matsubara axis. At a critical value of $\nu_0$, the two vortices merge on the Matsubara axis and at even smaller $\nu_0$ split along it, creating two zeros of $\Delta(\omega_m)$. We analyzed the gap function along the real axis and verified that each vortex in the upper frequency half-plane gives rise to a $2\pi$ variation of the phase of the gap function.

Both $\Delta' (\omega)$ and $\Delta^{''} (\omega)$ can be extracted from ARPES data sets, and we call for modern data analysis on electron-phonon superconductors to extract the phase variation over a wide range of frequencies to obtain information about dynamical vortices.

\begin{acknowledgments}
We thank A. Abanov, I. Mazin, Y. Wu, and S.-S Zhang for useful discussions. MHC acknowledges financial support from the Carlsberg foundation. The work by AVC was supported by the NSF DMR-1834856.
\end{acknowledgments}

\bibliography{bibliography_1}

\end{document}